# Identifying Human Mobility Patterns using Smart Card Data

**Oded Cats**

Department of Transport & Planning, Faculty of Civil Engineering and Geosciences, Delft University of Technology, Stevinweg 1, 2628 CN Delft, The Netherlands. Email: o.cats@tudelft.nl

**Abstract**

Human mobility is subject to collective dynamics that are the outcome of numerous individual choices. Smart card data which originated as a means of facilitating automated fare collection has emerged as an invaluable source for analysing human mobility patterns. A variety of clustering and segmentation techniques has been adopted and adapted for applications ranging from passenger demand market segmentation to the analysis of urban activity locations. In this paper we provide a systematic review of the state-of-the-art on clustering public transport users based on their temporal or spatial-temporal characteristics as well as studies that use the latter to characterise individual stations, lines or urban areas. Furthermore, a critical review of the literature reveals an important distinction between studies focusing on the intra-personal variability of travel patterns versus those concerned with the inter-personal variability of travel patterns. We synthesize the key analysis approaches and based on which identify and outline the following directions for further research: (i) predictions of passenger travel patterns; (ii) decision support for service planning and policy evaluation; (iii) enhanced geographical characterisation of users' travel patterns; (iv) from demand analytics towards behavioural analytics.

**Keywords**: Public transport; Smart card data; Market segmentation; Clustering; Urban analytics.

## 1. Introduction

Human mobility is subject to collective dynamics that are the outcome of numerous individual choices. Even though individual needs and travel choices exhibit large variability and the urban and regional areas in which those they are embedded are subject to great diversity, there is evidence to suggest that human mobility manifests some recurring features throughout history and across geographies (Ahmed and Stopher 2014). The growing availability of disaggregate mobility data enables the analysis of temporal and spatial patterns and the link between microscopic behaviour and resulting aggregate flows (Schläpfer et al. 2021). Geo-location traces are increasingly available from mobile phone and GPS data, social media posts and travel-related records such as automated fare collection (AFC) records (see Barbosa et al. 2018), giving rise to a plethora of studies analysing human mobility patterns and the decomposition thereof in the past decade. In this paper we provide a systematic review of the state-of-the-art on identifying human mobility patterns from smart card data and propose an outlook for addressing persisting challenges.

Smart card transactions offer a unique and rich source of passively collected data that enable the analysis of individual travel patterns. Pelletier et al. (2011) reviewed the technological advancements making the analysis of smart card data possible, followed by a review of strategic, tactical and operational applications of smart card data in public transport. Another relevant work is the one by Yue et al. (2014) who reviewed trajectory-based travel behaviour studies, where smart card data is covered as one of the emerging sources enabling the analysis of human mobility patterns, replacing travel diaries and stated preferences data. Their main focus is on the properties of different trajectory data categories. In a recent review, Hussain et al. (2021) examined recent developments in using smartcard data for the purpose of Origin-Destination matrix estimation. The latter - as well as related inference algorithms that are performed in order to estimate travel destinations, vehicles boarded, transfer locations and home-zones - is a pre-requisite for some of the research devoted to identifying human patterns based on smart card data.

In the last decade, an extensive research attention has been devoted to the identification and classification of mobility patterns from smart card data. Researchers have adopted and adapted a variety of clustering and segmentation techniques for a series of applications ranging from passenger demand market segmentation to the analysis of urban activity locations. The objective of this paper is to provide a systematic review of this body of knowledge, synthesize the key analysis approaches and offer a research agenda for addressing the remaining gaps.

Our search strategy was based on applying combinations of 'Smart card data' and one or more of the following keywords when performing the bibliometric search: 'Clustering', 'Travel patterns', 'Mobility patterns', 'User segmentation', 'Spatial', 'Temporal' and 'Spatiotemporal'. The databases of Scopus and Google Scholar were used to identify all relevant literature. The search, last performed on June 29, 2022, resulted in 698 papers of which 56 papers were identified as directly relevant for the scope of this paper. Some of those excluded performed aggregate statistical analysis of mobility patterns rather than identifying distinctive ones. Other papers excluded from this review focused on inference algorithms, modes other than public transport, aggregate analysis of statistics such as OD matrix estimation, ridership, service evaluation and travel time distributions, and the statistical distributions thereof.

All of the studies included in our analysis were published in the last decade, the vast majority (84%) of which in the last five years, from 2017 onwards. The selected papers were then categorized based on whether they analyse individual mobility patterns or network and urban analytics. Based on the papers reviewed, an additional distinction was made between studies that characterise individual mobility patterns based exclusively on their temporal patterns and studies that also consider spatial features as part of the pattern identification.

The remainder of the paper is organised as follows. The following section, Section 2, provides a synthesis of the literature concerned with identifying and characterising travellers' mobility patterns whereas Section 3 reviews studies that use the latter to characterise individual stations, lines or urban areas. We then offer a research agenda in Section 4, outlining key promising directions for future research.

## 2. Characterising Individual Mobility Patterns

Human mobility patterns are characterised by their temporal and spatial features. Most of the past work devoted to analysing individual mobility patterns using smart card data has focused on their temporal characteristics and is reviewed in the following sub-section 2.1. Thereafter, we turn in sub-section 2.2 into reviewing works that have explicitly considered both spatial and temporal aspects of user travel patterns as input to the market segmentation analysis.

### 2.1 Temporal patterns of individual users' travel

Table 1 provides a summary of the 22 studies analysing temporal user travel patterns, including their main aim, the variable(s) which are subject to analysis, the clustering technique employed, contextual information used in the analysis and in the interpretation of the results, and the key features of the case study application.

A critical review of the literature reveals an important distinction – yet only seldom made explicit by the studies themselves - between studies focusing on the *intra-personal variability* of travel patterns versus those concerned with the *inter-personal variability* of travel patterns. The former aims at classifying individual in terms of the stability of the temporal features of their travel patterns whereas the latter aims at segmenting users based on the characteristics of their regular temporal patterns.

Studies focusing on intra-personal stability have used various temporal features to analyse travel stability. The latter was investigated in terms of departure times (Kieu et al. 2015a, Manley et al. 2018), by the number of trips made on each day of the week (Deschaintres et al. 2019) or a combination thereof (Moradi and Trepanier 2020). A couple of studies have specifically looked into identifying so-called 'extreme' travellers defined by the percentage and frequency of trips in certain time periods or of certain duration (Long et al. 2016) or in terms of the number of days travelled per week, travel frequency per day of the week and temporal differences between daily trips (Cui and Long 2019).

Segmenting users based on their inter-personal variability can reveal similarities and differences in their temporal travel habits. Several studies have investigated aggregate indicators of travel frequency such as overall frequency of use and frequency of travel characteristics such as travelling by train and performing transfers (Kieu et al. 2018) or number of travel days (Egu and Bonnel 2020). Since weeks are considered to be the fundamental unit of recurrent travel schedules, the number of trips per day of the week has been often considered an important feature for market segmentation (Viallard et al. 2019), especially in combination with the starting hour (Briand et al. 2017, El Mahrsi et al. 2017, Liu and Cheng 2020, Cats and Ferranti 2022a), resulting in an hour-by-hour weekly travel profile per user which is then subject to clustering. User segmentation can then also be used for predicting the travel frequency per user per time-of-day as a conditional probability (Yang et al. 2018). Individual travel diaries allow to specifically focus on the timing of (certain) activities such as boarding times when travelling to and back from work (Ji et al. 2019), the start times as well as duration of activities (Medina 2018) and the activity sequence structure (Goulet-Langlois et al. 2016, Lei et al. 2020).

Smart card data facilitates the construction of individual travel diaries, which in turn enable the investigation of travel patterns amongst specific user groups. Many of the studied reviewed have analysed how the temporal patterns identified vary amongst users with different fare product types using the latter as a contextual post-analysis variable, some of which contain information on different user groups such as school pupils, higher education students, and retired or older users (Kieu et al. 2015a, Briand et al. 2017, El-Mahrsi et al. 2017, Deschaintres et al. 2019 and Egu and Bonnel 2020). A single study has specifically focused on a selected user group, namely on the variability of trip frequency per hour by older travellers (Liu et al. 2021). In several studies, personal information available from smart card registrations has also been used, for at least a sample of users included in the analysis (Goulet-Langlois et al. 2016, Liu et al. 2021) or from a matching household survey (Long et al. 2016). However, in most cases no socio-demographic data is available for individual card holders. Notwithstanding, by inferring card-holders place of residence from their travel patterns, one can link users to zonal socio-demographic characteristics available from census data such as ethnicity, employment and income (Liu and Cheng 2020, Cats and Ferranti 2022a).

Clustering or segmentation techniques for aim at identifying mutually exclusive and collectively exhaustive subsets of the population so as to maximize the intra-group similarities and minimize the inter-group similarities. The definition of the (dis)similarity metric is therefore crucial. In the context of user segmentation, the two most common clustering techniques are $k$-means (see Likas et al. 2003) and agglomerative hierarchical (see Rokach and Maimon 2005) clustering. The former finds the best partitioning of the dataset given a pre-defined number of clusters, $K$, with the distance metric pertaining to the centre of the cluster. A variant of which, $k$-medoids, selects an actual data point as a the center of each cluster. Agglomerative hierarchical techniques follow a bottom-up approach with iterative merging of groups of observations based on their similarity. While hierarchical approaches are more computationally expensive, they allow for greater variety of distance specifications and their output contains more information on relations between potential clusters and does not require the a-priori specification of the number of clusters. An additional technique used in several studies for clustering users based on the similarity between subsequent trips is DBSCAN (Khan et al. 2014) which relies on user-specified parameter values. A model-based approach such as Gaussian mixture model offer a latent profile assignment with probabilistic assignment of members to clusters which is especially attractive in the case of continuous variables (e.g. Briand et al. 2017).

A variety of clustering techniques have been employed in the analysis of temporal travel patterns. The most commonly used methods are: (i) K-means (Kieu et al. 2018, Yang et al. 2018, Deschaintres et al. 2019, Viallard et al. 2019, Liu and Cheng 2020, Liu et al. 2021, Eltved et al. 2021), (ii) DBSCAN (Medina 2018, Manley et al. 2018, Cui and Long 2019), (iii) hierarchical clustering (Goulet-Langlois et al. 2016, Ghaemi et al. 2017, He et al. 2018, Kieu et al. 2018, Deschaintres et al. 2019, Egu and Bonnel 2020, Moradi and Trepanier 2020, Cats and Ferranti 2022a) and (iv) multinomial or Gaussian mixture models (Briand et al. 2017, El Mahrsi et al. 2017, Ji et al. 2019, Cats and Ferranti 2022a). A single study applied a spatial affinity propagation technique (Kieu et al. 2018). Several studies have specifically focused on improving the projection of temporal distances to facilitate its hierarchical clustering (Ghaemi et al. 2017), on comparing the performance of alternative distance metrics when clustering users based on time-series data (He et al. 2018) or comparing alternative clustering techniques (Kieu et al. 2018).

Case study applications analysed smart card data from cities in Australia (Brisbane, Sydney), Canada (Gatineau, Montreal), China (Beijing, Nanjing, Shenzhen), Denmark (Copenhagen), France (Lyon, Rennes), Japan (Shizuoka), Singapore, Spain (Tarragona), Sweden (Stockholm) and United Kingdom (London). Most applications have considered metro systems, bus systems or a combination thereof, with few studies extending to other urban and suburban modes of public transport. Smart card data from most metro systems worldwide and bus systems from systems in East Asian analysed by studies included in this review contain both tap-in and tap-out data. Note that for most approaches taken in the analysis of temporal travel patterns no information regarding travel destination is per-se needed since boarding time (available from tap-in transactions) is the prime variable of interest used for constructing travel profiles that are then subject to clustering.

Few of the studies examining temporal travel patterns have also considered spatial elements such as the share of regular OD journeys (Kieu et al. 2015a) or information on travel modes and specific services (Manley et al. 2018). In addition, a number of studies have considered spatial attributes such as relating those to stations (Manley et al. 2018), inferred home location (Kieu et al. 2018, Liu and Cheng 2020, Cats and Ferranti 2022a), inferred home and job locations (Long et al. 2016), land use information (Lei et al. 2020), spatial entropy (Briand et al. 2017) as contextual variables.

Table 1: Summary of studies clustering user temporal travel patterns

| Study | Aim | Analysis | Clustering technique | Contextual information | Application | | |
|---|---|---|---|---|---|---|---|
| | | | | | City/Region | Modes | Tap-in (I) / Tap out (O) |
| Kieu et al. 2015a | Identify users that travel at regular times | Based on the intersection of % of regular OD journeys and % of habitual time journeys | A-priori segmentation of users (after DBSCAN of stops) | Journey attributes (travel time, mode, transfers), product type | South East Queensland, Australia | BCF | IO |
| Goulet-Langlois et al. 2016 | Identify users with similar activity sequence structure | Principal component analysis of multi-week activity sequences | Hierarchical | Socio-demographic information | London, United Kingdom | BMC | I – B; IO- MC |
| Long et al. 2016 | Identifying 'extreme' travellers | Percentage and frequency of trips in certain time periods or of certain length | NA | Household survey, inferred home and job locations | Beijing, China | BM | IO (for most trips) |
| Briand et al. 2017 | Year-on-year changes | Day-of-the-week and start hour per trip | Two-step Gaussian mixture model | Fare products, spatial entropy | Gatineau, Canada | B | I |
| El Mahrsi et al. 2017 | Identify groups with similar boarding times | Number of trips per hour-day | Multinomial mixture model | Fare products | Rennes, France | MB | I |
| Ghaemi et al. 2017 | Reduce high-dimensionality data by transforming the temporal use representation | Three-dimensional clock-like space projection of hour-day travel | Hierarchical | | Gatineau, Canada | B | I |
| He et al. 2018 | Assessing impacts of different distance metrics | Time-series | Hierarchical | | Gatineau, Canada | B | I |
| Kieu et al. 2018 | Scalability of spatial-behavioural segmentation | Frequency of use, frequency of train use, frequency of transfers, number of unique tap-in and tap-out pairs | Spatial affinity propagation, k-means, Hierarchical | Inferred home location, areas | New South Wales, Australia | BTC+ | IO |
| Manley et al. 2018 | Identifying regularities in individual travel patterns | Time series with binary indications for travel mode and specific services | DBSCAN | Stations | London, United Kingdom | BMC | I – B; IO- M |
| Medina 2018 | Common weekly patterns | Start time and duration of activities | DBSCAN | Household survey | Singapore | BTM | IO |
| Yang et al. 2018 | Predict travel frequency per user based on user segments | Entropy metric related to number of locations visited per time-of-day and weekday-weekend periods | k-means | | Shenzhen, China | M | IO |
| Cui and Long 2019 | Analysing intra-user regularity of travel patterns for 'extreme' and 'non-extreme' users | Number of days travelled per week, Travel frequency per day of the week and temporal differences between daily trips | A variant of DBSCAN (denominated OPTICS) | | Beijing, China | BM | IO (for most trips) |
| Deschaintres et al. 2019 | Intra-personal regularity of weekly patterns | Number of trips made on every day of the week | k-means for clustering usage-weeks, Hierarchical for clustering users | Fare products | Montreal, Canada | BM | I |
| Ji et al. 2019 | Determine the socio-economic and job-house | Commuting distance, boarding time to work | Gaussian mixture model | Household survey | Nanjing, China | M | IO |

| | | relation factors influencing metro commuting patterns | and boarding time from work | | | | | |
|---|---|---|---|---|---|---|---|---|
| Viallard et al. 2019 | Week-on-week evolution | Number of trips per day of the week | k-means | | Gatineau, Canada | B | I |
| Egu and Bonnel 2020 | Measuring interpersonal variability over days | Number of days on which both travellers have travelled | Hierarchical | Fare products | Lyon, France | BTM | I |
| Lei et al. 2020 | Temporal motifs for travel patterns | Construct a temporal graph of trip sequence, statistical analysis | NA | Land-use | Nanjing, China | M, Bike-sharing | IO (no inter-mode transfers) |
| Liu and Cheng 2020 | Extract travel patterns, enrich with demographics, illustrate for planning purposes (night tube) | Trip counts per day-of-the-week and start hour | k-means | Inferred home location, demographics | London, United Kingdom | MC | IO |
| Moradi and Trepanier 2020 | Measuring intra-user pattern stability | A binary vector of hour-day travel | Hierarchical (using a semicircle projection) | | Gatineau, Canada | B | I |
| Eltved et al. 2021 | User groups response to long-term disruptions | Regularity and intensity of travel as well as a weekday/weekend indicator | k-means | | Copenhagen, Denmark | MT | IO |
| Liu et al. 2021 | Role of age and environments on usage by older travellers | Entropy metric related to trip frequency per hour | k-means | Personal information available from card registration | Shizuoka, Japan | BC | IO |

| Cats and Ferranti 2022a | Identify market segments in terms of habitual temporal travel patterns | Number of journeys per day-of-the-week and hour-of-the-day combinations | Day of the week analysis: k-means followed by hierarchical; Hourly week pattern analysis: Gaussian Mixture Model | Inferred home location | Stockholm, Sweden | BTMC+ | I |

Modes: B- Bus; T – Tram; M- Metro; C – Commuter train; + - other modes (i.e. ferry, funiculars)
DBSCAN - Density-Based Scanning Algorithm with Noise; ISODATA - iterative self-organizing data analysis technique

## 2.2 Spatiotemporal patterns of individual users' travel

Table 2 provides a summary of user segmentation studies that have identified spatiotemporal travel patterns. Similarly to the main divide in the analysis of temporal travel patterns, also studies focusing on both spatial and temporal characteristics have either looked into intra-personal or inter-personal variability. The former considered in this context thus not only the extent to which users travel at the same times over the course of the analysis period (be it defined in terms of frequency per week, days of the week and/or time of the day) but also the locations between which they have travelled.

In an attempt to identify commuters, studies focusing on intra-personal variability have considered variables such as the number of active travel days and the similarity of first boarding times, stops and route sequences (Ma et al. 2013) or the prevalence of most frequent origin-destination and the respective boarding times and routes (Ma et al. 2017). A broader perspective, not limited to commuting and most commonly performed trips, considers the distribution of trips over daily time periods and the most often used ODs and stations (Kaewkluengklom et al. 2021), or adds greater detail in the form of the overall sequence of activity locations and their respective durations per day (Goulet-Langlois et al. 2017, He et al. 2021). A similar approach served the opposite goal in identifying irregular (undirected, potentially suspicious) travellers based on travel frequency per time window and station (Wang et al. 2019).

Studies focusing on inter-personal variability group users based on the characteristics of their regular spatiotemporal travel patterns (rather than the extent to which those are stable). Three studies have been identified in this category, all of which focusing on special user groups. Gutierrez et al. (2020) examined the combination of an array of temporal (travel frequency and number of travel days) and spatial (number of locations visited and routes used, prominence and identity of most visited locations) characteristics of tourists' travel patterns. Wang et al. (2019) investigated the activity spaces of vulnerable travellers' groups and employed to this end indicators related to both temporal (frequency, time entropy) and spatial (distance, radius and shape of activity space, place entropy) travel characteristics. Pieroni et al. (2021) focused on low-income workers in the context of a metropolis with large income gaps and contrasts their travel patterns with those of high-income ones. A snowball search identified one additional study which regressed the extent to which users tend to travel using the same route against user, OD and journey characteristics, albeit without seeking to identify patterns (Kim et al. 2017).

Two techniques are dominant when integrating spatial information into the clustering of individual users, k-means (Ma et al. 2013, Yang et al. 2018, Kaewkluengklom et al. 2021, Wang et al. 2019, Zhang et al. 2021a, Pieroni et al. 2021) and DBSCAN (Ma et al. 2013, Ma et al. 2017, Wang et al. 2019). Some exceptions exist for studies using a diverse array of travel features which use a Gaussian mixture model (Gutierrez et al. 2020) or hierarchical clustering (He et al. 2021), as well as a study identifying irregular users based on percentile cut-off of an entropy-based metric (Goulet-Langlois et al. 2017).

Of the ten studies identified which analysed the spatiotemporal characteristics of individual mobility patterns using smart card data, three of which have been conducted for the case of Beijing and the remaining, with one study each, for Brisbane, Gatineau, London, Sao Paulo, Shizuoka, Tarragona and Wuhu.

A special class of studies – not concerned with the identification of user segments, yet concerned with their individual travel patterns - is the one focusing on physical encounters and thereby contact networks and their potential consequences, in particular in the context of virus spreading. All these works construct a contact network based on the inferred passenger trajectories. Sun et al. (2013)

analysed the statistical properties of the topological indicators – including the clustering coefficient – of the resulting physical encountering network for bus users in Singapore. Liu et al. (2020) searched for the co-existence of passengers at stations and on-board vehicles using DBSCAN. Their application considered the metro networks of Shenzhen and they analyse the frequency and duration of encounters among pairs of passengers. Qian et al. (2021) analysed the dynamics of virus spreading in contact networks and the effectiveness of control strategies, namely vaccination and quarantine, using contract networks constructed from the metro systems of Guangzhou, Shanghai and Shenzhen in China.

It is evident that the number of studies that consider the spatial characteristics of travel patterns in the identification of user segments is comparatively small compared with the body of literature devoted to temporal aspects of human travel using smart card data. Furthermore, all of the relevant studies identified in this review have analysed spatial features in conjunction with temporal ones, rather than analysing spatial aspects in isolation. In line with the scope of this review, studies that analysed the spatiotemporal characteristics of ridership flows or origin-destination matrices are not included in this analysis. The limited research effort devoted to the consideration of spatial features as part of user segmentation arguably stems from its greater complexity as compared to temporal aspects. The latter can easily be discretized and are universal and therefore the transferability of the associated feature definition and clustering techniques is fairly straightforward. In contrast, spatial features are subject to local variations and are often labelled, their discretization is not trivial and the findings cannot be easily made transferable to other contexts. Notwithstanding, there is much room for further research in this domain as elaborated in Section 4.

Table 2: Summary of studies clustering user spatial-temporal travel patterns

| Study | Aim | Analysis | Clustering technique | Contextual information | Application | | |
|---|---|---|---|---|---|---|---|
| | | | | | City/Region | Modes | Tap- in (I) / Tap out (O) |
| Ma et al. 2013 | Distinguish between travellers based on the regularity of spatiotemporal characteristics | Spatial and temporal features such as number of travel days, number of similar first boarding times and similar stops and route sequences | DBSCAN per user and K-means++ to cluster users | | Beijing, China | BM | IO (for most trips) |
| Goulet-Langlois et al. 2017 | Measuring regularity of individual travel patterns in terms of frequency and sequence | Sequence of activity locations and respective activity durations per day | NA (regular and irregular users are identified based on percentile values of an entropy metric) | | London, United Kingdom | BMC | I – B; IO- M |
| Ma et al. 2017 | Identifying commuter/ regular travellers | Number of days travelled, number of occurrences for most frequent start location, destination location and | DBSCAN followed by ISODATA | Online survey, areas between ring-roads | Beijing, China | BM | IO (for most trips) |

| | | respective boarding times and routes | | | | | |
|---|---|---|---|---|---|---|---|
| Wang et al. 2019 | Identify irregular (undirected) travellers | Weighted sum of travel frequency at visited stops and time periods | k-means for temporal patterns followed by DBSCAN for detecting irregular travellers | Stations | Beijing, China | M | IO |
| Gutierrez et al. 2020 | Latent profile analysis of tourists travel patterns | Variables related to number of active days, number of transactions, average group size, number of stops, municipalities and zones visited, main stops visited, number of routes used, and share of transactions at main stops and areas. | Gaussian mixture model | Stations | Tarragona, Spain | B | I |
| He et al. 2021 | Integrate spatial- and temporal aspects of daily behaviour | Dynamic time warping distance of stop-sequences in daily profile | Hierarchical clustering (with sampling) | | Gatineau, Canada | B | I |

| Kaewkluengklom et al. 2021 | Year-by-year changes in cluster membership | Distribution of trips over four daily time periods, top three most used ODs and most used stations | k-means | Card type, land-use type and density, population density | Shizuoka, Japan | BC | IO |
|---|---|---|---|---|---|---|---|
| Pieroni et al. 2021 | Identify the travel patterns of low-income workers | Nine variables pertaining to the four categories of temporal patterns and variability, spatial patterns and variability, activity pattern and variability, and socio-economic. | k-means | Land-use types | Sao Paulo, Brazil | BMC | I |
| Zhang et al. 2021a | Examine variations in activity spaces of vulnerable travellers' groups | Indicators related to travel frequency, travel distance, radius and shape index of activity space, and place and time entropy to reflect diversity of activities per month | k-means | Population groups based on card type (low-income, disabled and elderly) | Wuhu, China | B | I |
| Faroqui et al. 2022 | Identify targeted groups of passengers based | Similarity in terms of location, time | Agglomerative hierarchical | Activity location type (stop, vehicle, billboard) | South East Queensland, Australia | BCF | IO |

| | on their activities and trips for advertising purposes | and type of activity | technique (Ward method) | and purpose of ads (work, education, shopping) | | | |
|---|---|---|---|---|---|---|---|

## 3. Spatiotemporal Travel Patterns for Network and Urban Analytics

In the previous section we reviewed the literature on clustering public transport users based on the temporal or spatial-temporal characteristics of their individual travel patterns. Another, related, body of literature has sought to identify and characterise clusters of (public transport) network elements – most often stops/stations but in a few cases service lines - based on the spatial-temporal characteristics of their respective users, as well as the urban area in which the networks are embedded.

Table 3 presents a summary of studies that performed network and urban analytics based on spatiotemporal travel patterns derived from passenger data. The objectives of the 24 studies included in this review vary greatly. Pioneering studies in the field sought to identify key activity centres using passenger flow data (Roth et al. 2011, Kim et al. 2014, Cats et al. 2015). Those have been followed by a series of studies that investigated the relation between the results of the station clustering and land-use patterns and places of interest in proximity to these stations or characteristics of the respective zones (Kim et al. 2017, Zhang et al. 2018, Zhao et al. 2019, Gan et al. 2020, Kim 2020, Zhuang et al. 2020). Recent studies have extended this to the analysis of how urban structure has evolved over time (Wang et al. 2021, Zhang et al. 2021b). Another stream of works has focused on the generation of origin-destination matrices by aggregating stops into demand zones (Kieu et al. 2015b, Luo et al. 2017), characterising the demand patterns associated with individual stations (Zhong et al. 2015, El-Mahrsi et al. 2017, Wang et al. 2017, Tang et al. 2018, Li et al. 2020, Zhou et al. 2022, Park et al. 2022), and clustering stations (Cats and Ferranti 2022b) or lines (Wang et al. 2020, Yap et al. 2019) that exhibit homogenous activity patterns.

With the exception of the latter group focusing on service lines which is aimed at supporting service planning, all other studies have used individual stations or set of stations in proximity (i.e. areas) as the unit of analysis subject to clustering. All of the abovementioned studies have utilized origin-destination flow information derived from smart card data in their analysis with the exception of one study relying on boarding volumes only (El Mahrsi et al. 2017) and two studies considering boarding and alighting flows (Cats et al. 2015, Gan et al. 2020), presumably due to data availability limitations.

Given the dynamic character of network and urban centre characteristics, it is not surprising that the vast majority of studies have considered the temporal properties of passenger flow distribution in their analysis. Most studies have integrated information on the temporal characteristics of the flows analysed as part of the clustering process (Cats et al. 2015, El Mahrsi et al. 2017, Kim et al. 2017, Tang et al. 2018, Zhao et al. 2019, Gan et al. 2020, Kim 2020, Li et al. 2020, Zhuang et al. 2020, Wang et al. 2021, Zhang et al. 2021b, Cats and Ferranti 2022b). In addition, several studies have investigated how the results of their clustering vary for different time periods (Luo et al. 2017, Wang et al. 2017, Yap et al. 2019, Wang et al. 2020).

Methods deployed for the clustering of stops (or sets of stops) and lines include k-means or variants thereof (Kim et al. 2017, Luo et al. 2017, Tang et al. 2018, Zhao et al. 2019, Gan et al. 2020, Cats and Ferranti 2022b, Zhou et al. 2022, Yong et al. 2021), variants of hierarchical methods (Roth et al. 2011, Kim et al. 2014, Cats et al. 2015, Kim 2020), DBSCAN (Kieu et al. 2015b, Yap et al. 2019), Poisson or Gaussian mixture models (El-Mahrsi et al. 2017, Wang et al. 2021), affinity propagation (Wang et al. 2017, Zhuang et al. 2020) and a discriminative functional mixture model (Park et al. 2022). In addition, a number of studies adopted community detection techniques (Zhong et al. 2015, Yap et al. 2019, Zhang et al. 2019, Wang et al. 2020, Zhang et al. 2021b). The latter – which have not been adopted by any of the studies performing user segmentation reported in sections 2.1 and 2.2 – is a graph-based technique which partitions a network into communities by grouping together nodes that are densely connected with those belonging to different communities being only sparsely connected, where the strength or distance of a connection is indicated by the selected link labelling.

Largely reflecting the geographical distribution of the research groups that have conducted related research, studies have analysed urban and system structure for cities from Australia (Brisbane), China (Beijing, Nanjing, Shanghai, Chongqing), France (Rennes), the Netherlands (The Hague), Singapore, South Korea (Seoul), Sweden (Stockholm) and the United Kingdom (London). The majority of studies have utilised (only) metro data whereas several studies included all public transport modes present in the case study area to best reflect the underlying travel demand patterns, including in areas which might be underserved by the metro.

Given the scope of this review, the discussion above and Table 3 are limited to studies that identify patterns in how passenger flows differ across the system for the purpose of either network or urban analytics. The bibliometric search yielded several adjacent studies that are either grounded in complex system theory or in urban planning. Studies within the realm of the former investigate statistical properties of travel flows across the network, such as the stability of how travel patterns vary over days for the networks of London, Singapore and Beijing (Zhong et al. 2016) or calculated several measures of urban diversity based on the temporal pattern of flows per spatial units (Sulis et al. 2018). Examples of the latter include the analysis of year-on-year change in the job to worker ratio per station in the Beijing metro network (Huang et al. 2019) and the identification of employment areas in Beijing and the evolution thereof based on temporal characteristics of incoming and outgoing flows (Huang et al. 2021). Note that these studies, unlike the studies included in Table 3, use descriptive statistics to observe spatial variations.

Table 3: Summary of studies clustering network elements based on related travel patterns

| Study | Aim | Analysis | Clustering technique | Contextual information | Application | | |
|---|---|---|---|---|---|---|---|
| | | | | | City/Region | Modes | Tap- in (I) / Tap out (O) |
| Roth et al. 2011 | Identify the polycentric structure of activity centers | Descending order to inflows and bundling stations in proximity | Hierarchical (own variant) | Travel distance | London, United Kingdom | M | IO |
| Kim et al. 2014 | Identify zones and related flows | Average frequency of flows | Hierarchical (own variant) | | Seoul, South Korea | M | IO |
| Cats et al. 2015 | Identifying and classifying activity centers | Temporal profile of metrics related to boarding and alighting flows | Two-stage hierarchical for identification and classification of geographical clusters | | Stockholm, Sweden | BTMC | I |
| Kieu et al. 2015b | Computational efficiency of travel pattern clustering | Based on OD pairs | DBSCAN | | South East Queensland, Australia | BC+ | IO |
| Zhong et al. 2015 | Variability of relations between stops over time | Group stops with high inter-flows | Community detection | | Singapore | BM | IO |
| El Mahrsi et al. 2017 | Identifying stations' roles based on temporal profile | Boarding per hour and day per station | Poisson mixture model | | Rennes, France | MB | I |

| Kim et al. 2017 | Relation between travel patterns and station area | Boarding and alighting per hour and day of the week | k-means | Land-use and socio-demographic variables | Seoul, South Korea | M | IO |
|---|---|---|---|---|---|---|---|
| Luo et al. 2017 | Identify clusters of stops with similar travel patterns for data-driven OD generation | Ratios between intra-cluster flows and inter-cluster flows and distances | k-means | Time of day and weekdays/ weekends | The Hague, the Netherlands | BT | IO |
| Wang et al. 2017 | Identify functions of stations using static geographical information | Origin and destination stations | Affinity propagation | Points of Interest (POI) characteristics; time of day and weekdays/ weekends | Shanghai, China | M | IO |
| Tang et al. 2018 | Identify temporal characteristics of stations | Boarding and alighting per time of day period | k-means |  | Shanghai, China | M | IO |
| Zhang et al. 2018 | Estimate the impact of zonal variables on the spatial distribution of demand and differences between PT and taxi | Average daily trips | Community detection | Population density and average income per zone | Singapore | BM, Taxi | IO |
| Yap et al. 2019 | Identify hubs and determining line bundles to be prioritised for transfer synchronization | Constructing a graph representing transfer flows between | DBSCAN for hub identification followed by modularity-based community | Time periods | The Hague, the Netherlands | BT | IO |

| | | neighbouring stops | detection for line bundles determination | | | | |
|---|---|---|---|---|---|---|---|
| Zhao et al. 2019 | Correlation analysis between clustering results and surrounding land uses | Normalized hourly ridership per station | k-medoids | Land uses | Nanjing, China | M | IO |
| Gan et al. 2020 | Explaining ridership profile with local environment variables | Normalized boarding and alighting per hour | k-means | Land-uses | Nanjing, China | M | IO |
| Kim 2020 | Enhance urban structure analysis with spatial mobility patterns (in addition to temporal) | Similarity between the distribution of inflow and outflow probabilities per time period | Spectral clustering followed by hierarchical clustering | Land-use, residential and employment density | Seoul, South Korea | MB | IO |
| Li et al. 2020 | Identify important features for classifying stations | PCA of eight indicators that relate to shares of passengers in different periods and statistics of the time series of passenger counts | Own 2-step heuristic | | Beijing, China | M | IO |
| Wang et al. 2020 | Identify the hierarchy amongst lines | Representing transfer flows | Community detection | Time periods | The Hague, the Netherlands | BT | IO |

| | | using a dual graph (C-space) | | | | | |
|---|---|---|---|---|---|---|---|
| Zhuang et al. 2020 | Relate stations to each other and location type | Boarding per time of day period and day | Affinity propagation | Points of Interest (POI) characteristics | Singapore | BTM | IO |
| Wang et al. 2021 | Identifying functional areas and their changes over years | Frequency of visits per station per time period | Gaussian mixture model | | Beijing, China | M | IO |
| Zhang et al. 2021b | Explaining changes in movements over years using station area variables | Weekday peak time flows | Community detection | Socio-economic data at neighbourhood-level (population and job density, median house price, planning policy) | London, United Kingdom | M | IO |
| Cats and Ferranti 2022b | Identify and classify urban areas based on how attractive they are as activity centers | Share of visits by non-residents (based on home-zone inference) from all visits for stations included in a clustered area per day-of-the-week and hour-of-the-day combination | k-means | Time periods | Stockholm, Sweden | BTMC+ | I |
| Park et al. 2022 | Forecast daily pattern of | Number of incoming | Discriminative functional | | Seoul, South Korea | M | IO |

| | incoming passengers | passengers per time window | mixture model (fun FEM) | | | | |
|---|---|---|---|---|---|---|---|
| Yong et al. 2021 | Identifying employment, residential and balanced station types | Job-housing ratio and commute properties | k-means | Residential and employment density | Chongqing, China | M | IO |
| Zhou et al. 2022 | Topological and ridership characteristics | Entrance, exit and passing flows as well as degree and betweenness centrality | k-means | | Beijing, China | M | IO |

## 4. Research agenda

As is evident from the synthesis of the literature provided in sections 2 and 3, there is an increasing body of research that has either characterised individual travel patterns or performed network and urban analytics based on the associated travel patterns. Based on the review of the literature several on-going trends in this research domain can be identified:

- *Specific user groups*. There is increasing interest in investigating travel patterns for particular target segments which exhibit distinctive travel patterns ranging from tourists (Gutierrez et al. 2020) and older travellers (Liu et al. 2021) to lower income individuals (Pieroni et al. 2021).
- *Enriching smart card data with socio-demographic data*. This has been accomplished by means of including personal information available from card registrations (Goulet-Langlois et al. 2016, Liu et al. 2021) or a matching household survey (Ling et al. 2016). Alternatively, zonal socio-demographic data was linked by inferring the home-zone location of card holders (Liu and Cheng 2020, Cats and Ferranti 2022a).
- *Evolution of travel patterns*. The availability of smart card data over the course of a substantial time period allows for the analysis of changes in observed travel patterns in terms of identified market segments (Briand et al. 2017, Viallard et al. 2019), station clusters (Huang et al. 2021) or underlying urban characteristics (Zhang et al. 2021b).

Based on our critical review of the state-of-the-art we identify in the following four knowledge gaps and outline related directions for further research.

- *Predictions of passenger travel patterns*. The availability of system-wide passenger data opens avenues for the development of schemes aimed at the prediction of passenger travel patterns. Such predictions can facilitate service design adjustments as well as supply- and demand-management measures to better cater or steer anticipated flows. A couple of studies included in this review already do so specifically for travel frequency per user (Yang et al. 2018) and the daily pattern of arriving passengers at stations (Park et al. 2022).

   The future development of prediction schemes can be structured along two axes: *short-term vs. long-term* and *disaggregate vs. aggregate*. The former distinction pertains to the time horizon for which the prediction is made, whether it concerns within-day downstream conditions measured in minutes to hours or considering a timespan of days or even months. Short-term predictions pose requirements on computational efforts and possibly also on the real-time availability of smart card data whereas long-term predictions can rely on offline applications. Several recently deployed fare collection systems enable the real-time availability of smart card data, albeit those are still far from common practice. The distinction between disaggregate vs. aggregate pertains, respectively, to whether the unit of analysis that is subject to prediction relates to trip characteristics of each individual traveller or to the travel pattern characteristics which are the outcome of collective dynamics, such as boarding, alighting and on-board passenger flows.

   All four combinations of short- and long-term, disaggregate and aggregate predictions are relevant as future research directions and can facilitate respective applications and interventions by service providers. Several pioneering studies proposed a logistic regression model for predicting the next trip to be performed by an individual (Zhao et al. 2018) and an elasticity model for short-term aggregate predictions (van Oort et al. 2019).

- *Decision support for service planning and policy evaluation.* The analysis of travel patterns should ultimately inform planners in making decisions ranging from strategic and tactical to operational planning. This can be further facilitated by fusing long- and short-term predictions of travel patterns. While most of the studied included in this review have alluded to potential relevant applications, none of them has made an explicit link to service planning tasks, with the exception of Yap et al. (2019) and its relevance for service coordination.

  There is thus clearly a missing link between the development and application of demand, urban and network clustering analysis on one hand and service design and decision support on the other hand, which is key for the value proposition of the literature reviewed here. Future research should bridge this gap by making explicit how the techniques and insights developed based on smart card data analytics can translate into service improvements. The latter can range from ex-ante policy assessment (see Kholodov et al. 2021 for an example concerned with price elasticities) and network and capacity allocation adjustments and even the design of flexible services (see Qiu et al. 2019 for an original contribution in this direction) on one hand to real-time management such as information provision and control measures on the other hand. In particular real-time contrl strategies will greatly benefit from advancements in short-term passenger flow predictions. Another promising direction is the development of fare products and fare schemes (see Halvorsen et al. 2020 for a relevant framework) based on the market segmentation performed using smart card data. For example, changes in travel patterns induced by the COVID pandemic crisis and work-from-home habits call for the development of new subscription models which can be devised to cater for changes in spatiotemporal user clusters.

- *Enhanced geographical characterisation of users' travel patterns.* The analysis of individual mobility patterns and the related user clustering and market segmentation have insofar largely focused on their temporal features (section 2.1) or the stability and variability of travel destinations (section 2.2). However, there is lack of knowledge on the geographical properties of individual travellers by characterising spatial features of locations visited based on disaggregate longitudinal data. For the latter to be made available it is essential to comply with rigid privacy protocols due to the sensitivity of sequential geo-location data. Moreover, the aforementioned trend of enriching smart card data with socio-demographic data has been observed only for temporal user clustering but is yet to enter studies focusing on spatial clustering. The analysis thereof will enable identifying the extent to which different user groups visit different parts of the urban area and when analysed in the aggregate, the composition of travellers visiting various zones and the extent to which various destinations are visited by people with different backgrounds. Community detection techniques may be apply for user segmentation with the goal of identifying groups of travellers that share similar geographical features of their mobility patterns.

  In the analysis of spatial characteristics it is advised to extract features that can be generalized to other contexts such as density, diversity and geographical scope of destinations visited rather than focus on labelled ones that pertain to local features (e.g. location names). Given the significant role that features of the local urban environment are likely to play in determining the spatial characteristics of traveller's mobility, there is a need to gain more knowledge on how those are manifested in diverse settings. While these variables vary also within any given city, they may be confounded and the range of values might be limited. A cross-city comparison will allow concluding on relevant determinants such as urban form, density, modal competition, land-use and demographics.

- *From demand analytics towards behavioural analytics.* While the identification and characterisation of human mobility patterns offers remarkable demand analytics insights it does not advance our knowledge on the underlying determinants of travellers' behaviour. A separate large body of research is devoted to the estimation of travel choice determinants from smart card data, primarily limited to the estimation of route choice models and the impact of crowding therein (e.g. Hörcher et al. 2017, Yap et al. 2020). There is great potential in marrying these two research streams so as to support the development of behavioural models for users with distinctive mobility patterns, for example by means of estimating latent class models that are informed by the user clustering results.

  Obtaining behavioural insights will also be stimulated by segmenting mobility patterns based on a combination of travel features that provide additional information on users experience. A couple of recent examples making advancements in this direction include the identification of central travellers for virus spreading by means of the simultaneous consideration of the distance travelled, the number of fellow travellers one has been exposed to and the degree of exploration (El Shoghri et al. 2020) and the neural network approach taken for the joint consideration of travellers' departure and arrival time, travel location, travel distance and point of interest at the destination (Li et al. 2021).

Smart card data which originated and has been designed as a means of facilitating automated fare collection has clearly emerged as an invaluable source for analysing human mobility patterns. Fusing smart card data with other sources of geo-location data will potentially allow analysts to consider additional aspects of human mobility beyond those performed by means of passenger transportation. Moreover, for the research analysing mobility patterns using smart card data to become an integral part of service design and decision making and offer insights that go beyond describing current travel patterns and related user segments, future research must combine methodological advancements in data science, machine learning and clustering techniques with knowledge in travel behaviour, demand forecasting and transport planning.